\documentclass[prl,twocolumn,showpacs,preprintnumbers,amsmath,amssymb,superscriptaddress]{revtex4}
\usepackage{graphicx}
\usepackage{dcolumn}
\usepackage{bm}

\begin{document}

\title{Graphite from the viewpoint of Landau level spectroscopy:\\
An effective graphene bilayer and monolayer}
\author{M. Orlita}
\email{orlita@karlov.mff.cuni.cz} \affiliation{Grenoble High Magnetic Field
Laboratory, CNRS, BP 166, F-38042 Grenoble Cedex 09, France}
\affiliation{Institute of Physics, Charles University, Ke Karlovu 5, CZ-121~16
Praha 2, Czech Republic} \affiliation{Institute of Physics, v.v.i., ASCR,
Cukrovarnick\'{a} 10, CZ-162 53 Praha 6, Czech Republic}
\author{C. Faugeras}
\affiliation{Grenoble High Magnetic Field Laboratory, CNRS, BP 166, F-38042
Grenoble Cedex 09, France}
\author{J. M. Schneider}
\affiliation{Grenoble High Magnetic Field Laboratory, CNRS, BP 166, F-38042
Grenoble Cedex 09, France}
\author{G. Martinez}
\affiliation{Grenoble High Magnetic Field Laboratory, CNRS, BP 166, F-38042
Grenoble Cedex 09, France}
\author{D. K. Maude}
\affiliation{Grenoble High Magnetic Field Laboratory, CNRS, BP 166, F-38042
Grenoble Cedex 09, France}
\author{M. Potemski}
\affiliation{Grenoble High Magnetic Field Laboratory, CNRS, BP 166, F-38042
Grenoble Cedex 09, France}
\date{\today}

\begin{abstract}
We describe an infrared transmission study of a thin layer of bulk graphite in magnetic fields up to $B=34$~T. Two
series of absorption lines whose energy scales as $\sqrt{B}$ and $B$ are present in the spectra and identified as
contributions of massless holes at the $H$ point and massive electrons in the vicinity of the $K$ point, respectively.
We find that the optical response of the $K$ point electrons corresponds, over a wide range of energy and magnetic
field, to a graphene bilayer with an effective inter-layer coupling $2\gamma_1$, twice the value for a real graphene
bilayer, which reflects the crystal ordering of bulk graphite along the $c$ axis. The $K$ point electrons thus behave
as massive Dirac fermions with a mass enhanced twice in comparison to a true graphene bilayer.
\end{abstract}


\maketitle

Recent interest in graphene~\cite{GeimNatureMaterial07,CastroNetoRMP08}, a
truly two-dimensional system with its simple, but nevertheless, for solids,
unconventional electronic states, has focused attention on the properties of
Dirac-like fermions in condensed matter physics in general. Whereas,
two-dimensional massless Dirac
fermions~\cite{NovoselovNature05,ZhangNature05,BergerScience06}, characteristic
of graphene have been widely investigated, far fewer experiments have been
devoted to massive Dirac fermions which are specific to a graphene
bilayer~\cite{NovoselovNaturePhys06,HenriksenPRL08}, which represents a further
example of a two-dimensional system with a highly unusual band
structure~\cite{McCannPRL06}. Perhaps surprisingly, Dirac dispersion relations
can also be found in graphite, a three dimensional, bulk material which
consists of Bernal-stacked weakly coupled graphene layers.

The standard Slonczewski-Weiss-McClure (SWM) model of electronic states in
graphite~\cite{SlonczewskiPR58,McClurePR57} predicts a complex form for the
in-plane dispersion relation which changes considerably depending upon the
value of the momentum $k_z$ in the direction perpendicular to the layers.
Intriguingly, the SWM model predicts that in the vicinity of the $H$ point
($k_z=0.5$) the in-plane dispersion is linear and thus resembles a Dirac cone.
Such a dispersion has indeed been found in angle resolved photoemission
spectroscopy~\cite{ZhouNatPhys06,GruneisPRL08}, tunneling
spectroscopy~\cite{LiNatPhys07,LatyshevJPCS08}, as well as in Landau level
(LL)-spectroscopy~\cite{OrlitaPRL08}. The latter experiments, mainly focused on
transitions between LLs whose energy scales as $\sqrt{B}$, are generally
believed to exhibit far richer spectra in comparison to true
graphene~\cite{SadowskiPRL06,JiangPRL07,DeaconPRB07}, reflecting the inherent
complexity of the SWM model which includes no fewer than seven
parameters~\cite{ToyPRB77,LiPRB06}.

We show in this Letter that infrared magneto-absorption spectra of graphite, measured over a wide range of the
energy and magnetic field, can be interpreted in a very simple, transparent and elegant manner. Our results confirm, in
agreement with theoretical considerations~\cite{KoshinoPRB08}, that graphite can be viewed as an effective graphene
monolayer and bilayer. This theoretical picture is derived using a drastically simplified SWM model, which includes
only two parameters $\gamma_0$ and $\gamma_1$, describing the intra- and inter-layer tunneling respectively. In this
simplified picture, the dominant contribution to the optical response is provided by the $H$ point, where electron
states closely resemble graphene but with an additional double degeneracy, and by the $K$ point, where the energy
spectrum resembles a graphene bilayer, but with an effective coupling of $2\gamma_1$, twice enhanced compared to a real
bilayer system.

Remarkably, using this simple graphene monolayer plus bilayer view of graphite, we are able to correctly reproduce the
magnetic field evolution of all observed inter-LL transitions using only the SWM parameters $\gamma_0$ and $\gamma_1$,
with values which perfectly match those derived from studies of real graphene monolayer and bilayer systems.
Interestingly, the electronic states at $K$ point of graphite are found to mimic those of the graphene bilayer, but
with a doubled value of the effective mass, so that they might be useful to further explore the interesting physics of
massive Dirac fermions.

\begin{figure}
\scalebox{1.3}{\includegraphics{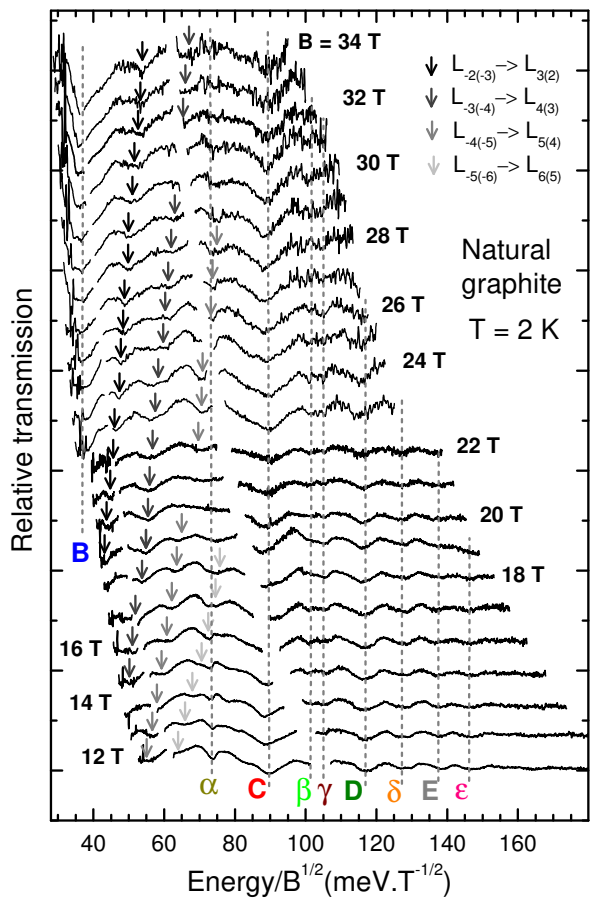}} \caption{\label{SPKT} (color online) Transmission spectra of a thin graphite
layer as a function of the magnetic field in the interval $B=12-34$~T. The plotted energy is scaled as $E/\sqrt{B}$ to
emphasize the Dirac fermion-like features in spectra (indicated by dashed vertical lines). Arrows denote transitions
arising at the $K$ point which evolve (nearly linearly) with $B$. Data above $B=23$~T were taken on second sample with
a higher density of graphite flakes.}
\end{figure}

\begin{figure*}
    \begin{minipage}{0.62\linewidth}
      \scalebox{1.15}{\includegraphics{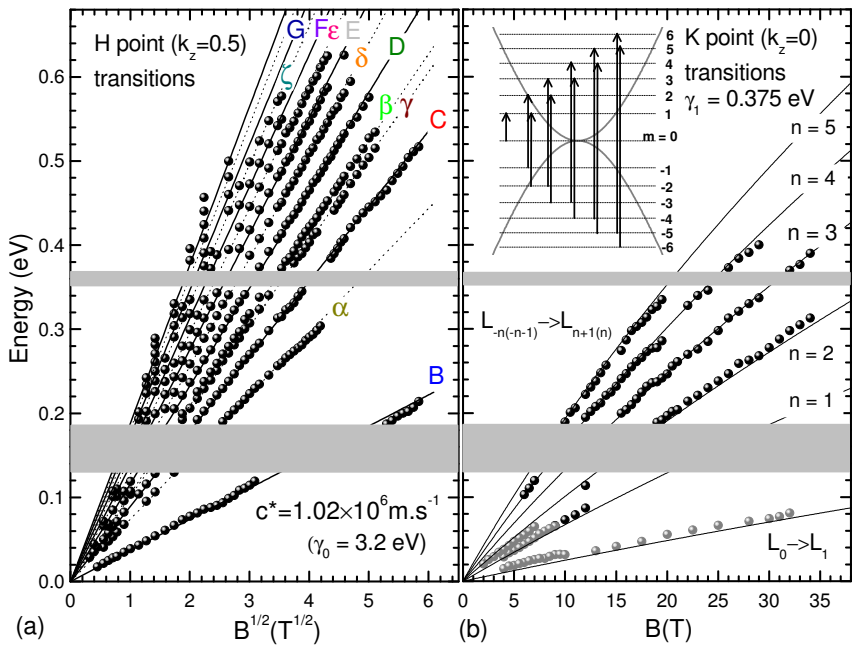}}
    \end{minipage}\hfill
    \begin{minipage}{0.34\linewidth}
      \caption{\label{FanChart} (color
online) (a): Positions of the absorption lines related to the $H$ point as a
function of $\sqrt{B}$. The solid and dashed lines represent expected positions
of absorption lines for $\tilde{c}=1.02\times 10^6$~m/s ($\gamma_0=3.2$~eV).
(b): $K$ point related absorption lines as a function of $B$. The solid lines
show expected dipole allowed transitions in a graphene bilayer with an
effective coupling $2\gamma_1$ calculated using Eq.~\eqref{Bilayer} for
$\gamma_0=3.2$~eV and $\gamma_1=0.375$~eV. Grey data points were taken on
highly oriented pyrolytic graphite, which exhibit a behavior nearly identical
to natural graphite~\cite{OrlitaJPCM08}. The inset schematically shows the
observed inter-band transitions in the effective bilayer.}
    \end{minipage}
\end{figure*}

Thin samples for the transmission measurements were prepared by exfoliation of a natural graphite crystal as described
in Ref.~\cite{OrlitaPRL08}. Some data is also presented for highly oriented pyrolytic graphite, which shows practically
identical, although slightly less pronounced features, in the magneto-transmission spectra~\cite{OrlitaJPCM08}. All
experiments were carried out on macroscopic, roughly circular-shaped samples, of several millimeters in diameter.
Measurements were performed in the Faraday configuration with the magnetic field applied along the $c$-axis of the
graphite. All spectra were taken with non-polarized light. To measure the magneto-transmittance of the sample in the
spectral range of 10-700~meV, the radiation of globar was delivered via light-pipe optics to the sample. The radiation,
detected by a Si bolometer, placed directly behind the sample and cooled down to a temperature of 2~K, was analyzed by
a Fourier transform spectrometer~\cite{SadowskiPRL06,OrlitaPRL08}. The transmission spectra were normalized by the
transmission of the tape and by the zero-field transmission, thus correcting for any magnetic field induced variations
in the response of the bolometer. The missing parts of the transmission spectra, indicated by grey areas in
Figs.~\ref{SPKT} and \ref{FanChart}, correspond to the spectral ranges where the tape is completely opaque.

Prior to presenting our experimental results, we outline a simple model of bulk graphite based on SWM
model~\cite{SlonczewskiPR58,McClurePR57}. Whereas the standard SWM model has seven tight-binding  parameters
$\gamma_0,\ldots,\gamma_5$, we limit ourselves here to only the most important hopping integrals $\gamma_0$ and
$\gamma_1$. In other words, we consider only the parameters which are relevant for the nature of the band structure in
a graphene monolayer and bilayer. In graphene, the intra-layer coupling parameter $\gamma_0$ is directly related to the
Fermi velocity, $\tilde{c}=\sqrt{3}a_0\gamma_0/(2\hbar)$, where the atomic distance is
$a_0=0.246$~nm~\cite{ChungJMS02},
and in a graphene bilayer, the inter-layer coupling $\gamma_1$ gives an estimate for the mass of the charge
carriers, $m=\gamma_1/(2\tilde{c}^2)$.

Within our simplified approach, we calculate the band structure along the $H-K-H$ line of the Brillouin zone, i.e. for
$-0.5<k_z<0.5$, which is essential for electrical and optical properties of bulk graphite. For the in-plane dispersion
of charge carries, we find~\cite{McClurePR57,SlonczewskiPR58,NakaoJPSJ76,ToyPRB77} that it has, for a given momentum
$k_z$, the form of a graphene bilayer with an effective coupling $\lambda\gamma_1$, where $\lambda=2\cos(\pi
k_z)$~\cite{NakaoJPSJ76,ToyPRB77}.

In a magnetic field, we obtain the LL spectrum for each effective bilayer, i.e. for each momentum $k_z$:
\begin{multline}\label{Bilayer}
\varepsilon^{\pm}_{n,\mu}=\pm\frac{1}{\sqrt{2}}\left[(\lambda\gamma_1)^2+(2n+1)E_1^2+\phantom{\sqrt{X}}\right.\\
\left.\mu\sqrt{(\lambda\gamma_1)^4
+2(2n+1)E_1^2(\lambda\gamma_1)^2+E_1^4}\right]^{1/2},
\end{multline}
where $\pm$ sign labels the electron(+) and hole(-) levels. LLs related to the touching electronic bands are obtained
for $\mu=-1$ and those related to bands split-off in energy by an amount $\pm\lambda\gamma_1$ are represented by
$\mu=1$. The touching bands can be in the parabolic approximation characterized by the mass
$m=\lambda\gamma_1/(2\tilde{c}^2)$. The in-plane coupling $\gamma_0$ enters Eq.~\eqref{Bilayer} via the Fermi velocity
$\tilde{c}$ and directly influences the characteristic  spacing of levels $E_1=\tilde{c}\sqrt{2e\hbar B}$. Note, that
our approach is a special case of the model used by Koshino and Ando~\cite{KoshinoPRB08}, who in an analogous way
calculated the spectrum of multilayer Bernal-stacked graphene with an arbitrary number of layers in an external
magnetic field.

The joint density of states (initial and final states), essential in our
magneto-optical experiments, has in the full as well as in our reduced SWM
model singularities  at two distinct points of the Brillouin zone, at the $K$
($k_z=0$) and $H$ ($k_z=0.5$) points, where electrons and holes are located,
respectively. Hence, the magneto-optical response of bulk graphite should be
governed by transitions between LLs defined by Eq.~\eqref{Bilayer} for $\lambda
\rightarrow 2$ ($K$ point) and $\lambda \rightarrow 0$ ($H$ point). Notably,
there is no singularity for $\lambda=1$, which corresponds to a real graphene
bilayer. Consequently, bulk graphite should, in magneto-optical experiments,
behave as a combination of a graphene bilayer with the effective coupling
$2\gamma_1$ and of a graphene monolayer, but with a twofold degeneracy
$\varepsilon^{\pm}_{n,-1}=\varepsilon^{\pm}_{n+1,1}$, in addition to the usual
twofold spin and valley degeneracies. The expected magneto-optical response of
bulk graphite should therefore contain hole-related features whose energy
evolves linearly with $\sqrt{B}$ originating in the vicinity of the $H$ point
together with absorption lines whose energy evolves roughly linear with $B$
corresponding to electrons at the $K$ point.

The magneto-transmission spectra taken for magnetic fields $B=12-34$~T on a thin layer of bulk graphite are presented
in Fig.~\ref{SPKT}. The transmission is plotted as a function of energy divided by $\sqrt{B}$ to facilitate the
identification of spectral features originating from around the $H$ and $K$ points. Plotted in this way, the
transitions denoted by Roman and Greek letters, do not shift for spectra recorded at different magnetic fields. Thus
they scale linearly with $\sqrt{B}$, see Fig.~\ref{FanChart}a, and are related to the $H$ point. The second set of
lines marked by vertical arrows shift with magnetic field and actually follow a nearly linear dependence with $B$, as
can be seen in Fig.~\ref{FanChart}b.

The transitions following a $\sqrt{B}$ dependence, corresponding to massless
holes around the $H$ point, have been thoroughly analyzed in our previous
work~\cite{OrlitaPRL08,OrlitaJPCM08}. Whereas, the absorption lines denoted by
Roman letters have their direct counterpart in spectra of true
graphene~\cite{SadowskiPRL06,JiangPRL07,DeaconPRB07,SadowskiSSC07,PlochockaPRL08,OrlitaPRL08II},
the Greek lines are in principle dipole forbidden in a pure 2D system of Dirac
fermions. Nevertheless, these transitions can be consistently explained  with
the same selection rule $\Delta n=\pm 1$, when the twofold degeneracy of LLs,
$\varepsilon^{\pm}_{n,-1}=\varepsilon^{\pm}_{n+1,1}$, at the $H$ point of bulk
graphite is properly considered. The Fermi velocity is extracted to be
$\tilde{c}=(1.02\pm0.02)\times10^6$~m/s, giving a rather precise measure of the
in-plane hopping integral $\gamma_0=(3.20\pm0.06)$~eV in bulk graphite, which
is the only parameter needed to describe all absorption lines originating at
the $H$ point.

Here we focus on transitions denoted by arrows in Fig.~\ref{SPKT}, whose energy
evolves nearly linearly with $B$. Taking into account the selection rule
$\Delta n=\pm 1$ and using the in-plane coupling estimated above to
$\gamma_0=3.2$~eV, we can interpret all the absorption lines in
Fig.~\ref{FanChart}b as dipole-allowed transitions in a graphene bilayer with
an effective coupling $2\gamma_1$ with $\gamma_1=(375\pm10)$~meV. The deduced
value for $\gamma_1$ is in a very good agreement with results obtained for a
real graphene bilayer~\cite{ZhangPRB08,KuzmenkoCM08} as well as on bulk
graphite~\cite{GruneisPRB08,KuzmenkoPRL08}. Our data  follow well the
theoretical predictions up to the highest energies, in contrast to the the
deviation reported for a real graphene bilayer~\cite{HenriksenPRL08}. Hence,
the electrons in the vicinity of the $K$ point can be described with a
reasonable accuracy using the model of a graphene bilayer, but with coupling
strength twice enhanced as compared to a true bilayer. The strength of the
coupling $2\gamma_1$ directly reflects the long-range Bernal stacking of
graphite along the $c$ axis. The twice enhanced coupling in the effective
bilayer can be understood using the example of semiconductor superlattices,
which are three-dimensional but nevertheless strongly anisotropic systems
resembling in some aspects the band structure of bulk graphite. The energy
difference, $\Delta_{SAS}$, between the bonding and anti-bonding state in a
symmetric double quantum well is simply half of the miniband width of the
superlattice created from the same wells~\cite{GoncharukPRB05}.

Our results also show that the parabolic approximation which is widely used for
the touching electronic bands in a bilayer and which directly leads to LLs
whose energy evolves linearly with magnetic field:
$\varepsilon^{\pm}_{n,-}\approx
\pm\hbar\omega_c\sqrt{n(n+1)}$~\cite{McCannPRL06,LiNatPhys07} is a good
approximation only in the vicinity of the charge neutrality point. As can be
seen in Fig.~\ref{FanChart}(b), the small deviation from a linear dependence,
predicted at higher energies by the simplified SWM model for an effective
bilayer, is reproduced in our data. The bilayer character of $K$ point
electrons also explains the non-linear evolution with $B$ of the
magneto-reflection spectra published recently~\cite{LiPRB06}. Nevertheless,
within the parabolic approximation, we obtain an effective mass of $K$ point
electrons in graphite of $m=\gamma_1/\tilde{c}^2\approx 0.063m_0$ in good
agreement with other cyclotron resonance experiments~\cite{Brandt88}. This mass
is a factor of two higher when compared to a true
bilayer~\cite{HenriksenPRL08}.

While, only two tight-binding parameters $\gamma_0$ and $\gamma_1$ are required to obtain a reasonable description of
the magneto-optical response of the $K$ point electrons over a wide range of energy and magnetic field, the influence
of the remaining hopping integrals $\gamma_2,\ldots,\gamma_5$, merits some consideration. In general, additional
coupling parameters should lift the electron-hole symmetry of the bilayer and the trigonal warping ($\gamma_3$) should
lead to a mixing of LLs, which in turn can give rise to additional dipole-allowed transitions.

The electron-hole asymmetry at the $K$ point should result in an energy difference of the transitions
L$_{-n}\rightarrow$L$_{n+1}$ and L$_{-n-1}\rightarrow$L$_{n}$. Indeed, some evidence for this splitting is visible in
the spectra, see e.g. the transition $n=3$ above $B\approx 20$~T in Fig.~\ref{SPKT}. Nevertheless, this effect is
relatively weak, comparable to the width of the absorption lines. The electron-hole asymmetry at the $K$ point of
graphite seems to be somewhat weaker in comparison to the observed asymmetry in a true graphene
bilayer~\cite{HenriksenPRL08,LiPRL09,KuzmenkoCM08}.

The influence of the trigonal warping $\gamma_3$ on the magneto-optical
response of the bilayer has been discussed by Abergel and
Fal'ko~\cite{AbergelPRB07}. They conclude that this parameter becomes important
only in the limit of low magnetic fields giving rise to a completely new set of
dipole-allowed transitions. Similar reasoning~\cite{NozieresPR58} explains the
observation of numerous harmonics of the cyclotron resonance of electrons in
bulk graphite at low magnetic fields~\cite{GaltPR56,DoezemaPRB79}.

Our simplified model also neglects the hopping integral $\gamma_2$ which is directly responsible for the semi-metallic
nature of bulk graphite. This parameter leads to a finite width of the doubly degenerate $E_3$ and consequently, the
$k_z$-dependent LLs $n=0$ and $n=-1$~\cite{NakaoJPSJ76}. Assuming a finite and negative value of
$\gamma_2$~\cite{Brandt88}, the $K$ point transition L$_{-1}\rightarrow$L$_0$ cannot be observed at any magnetic field
at low temperatures, as the Fermi level remains close to the middle of $n=0$ and $n=-1$ Landau bands even in the
quantum limit in graphite.

To conclude, the magneto-optical response of bulk graphite can, over a wide range of energy and magnetic field, be
understood within a picture of an effective graphene monolayer and an effective bilayer with a coupling strength
enhanced twice in comparison to a true graphene bilayer. This finding is in excellent agreement with predictions of a
drastically reduced SWM model which retains only two tight-binding parameters, namely the in-plane and inter-plane
coupling constants $\gamma_0$ and $\gamma_1$. It should be noted, that as the validity of the model is limited in the
vicinity of the Fermi level, it is not useful, for example, for the interpretation of magneto-transport experiments.
Nevertheless, bulk graphite remains a material of choice to study magneto-optical phenomena in systems with both
massless as well as massive Dirac fermions.

\begin{acknowledgments}
Part of this work has been supported by EuroMagNET II under the EU contract, by the French-Czech Project Barrande
No.~19535NF, by contract ANR-06-NANO-019 and by projects MSM0021620834 and KAN40010065.
\end{acknowledgments}


\end{document}